\documentclass[pra,aps,footinbib,twocolumn,superscriptaddress,citeautoscript,longbibliography]{revtex4-1}

\usepackage[utf8x]{inputenc}
\usepackage{array}
\usepackage{amssymb}
\usepackage{amsmath}
\usepackage{amsfonts}
\usepackage{color}
\usepackage{graphicx}
\usepackage{bm}
\usepackage{dsfont}
\usepackage{url}
\usepackage{braket}
\usepackage{empheq}
\usepackage{tikz}
\usepackage[unicode]{hyperref}
\usepackage{color}
\usepackage{soul}

\hypersetup{
   unicode=true,          
   plainpages=false,
   colorlinks=true,       
   citecolor=blue,        
}
\allowdisplaybreaks
\graphicspath{{.}{./fig/}}

\newcommand{\rme}{\mathrm{e}}
\newcommand{\rmi}{\mathrm{i}}

\newcommand{\x}{\mathrm{x}}
\newcommand{\y}{\mathrm{y}}

\newcommand{\n}{{\textbf{n}}}
\newcommand{\tn}{{\tilde{\textbf{n}}}}
\newcommand{\tm}{{\tilde{\textbf{m}}}}

\newcommand{\m}{{\textbf{m}}}
\newcommand{\lket}{\ket{\boldsymbol{\lambda}}}

\renewcommand{\d}{\ensuremath{\mathrm{d}}}

\newcommand{\kbar}{\mathchar'26\mkern-9mu k}

\newcommand{\HF}{\hat{H}_\mathrm{F}}
\newcommand{\F}{\hat{F}}

\newcommand{\Vt}{U(\hat{\boldsymbol{\theta}})}
\newcommand{\Vts}{U(\boldsymbol{\theta})}
\newcommand{\K}{\hat{K}}
\newcommand{\kap}{\hat{\kappa}}
\newcommand{\Keff}{\chi}

\newcommand{\sect}[1]{\section{#1} }

\newcommand{\pcsadd}{Center for Theoretical Physics of Complex Systems, Institute for Basic Science (IBS), Daejeon 34126, Republic of Korea}

\begin{document}

\title{Many-body dynamical localisation of coupled quantum kicked rotors}
\author{L. A. Toikka}
\email{lauri.toikka@gmail.com}
\affiliation{Institute for Theoretical Physics, University of Innsbruck, A-6020 Innsbruck, Austria}
\affiliation{\pcsadd}
\author{A. Andreanov}
\affiliation{\pcsadd}
\date{\today}

\begin{abstract}
The quantum motion of $N$ coupled kicked rotors is mapped to an interacting $N$-particle Anderson-Aubry-Andr\'e tight-binding problem supporting many-body localised (MBL) phases. Interactions in configuration space are known to be insufficient for destroying Anderson localisation in a system in the MBL phase. The mapping we establish here predicts that a similar effect takes place in momentum space and determines the quantum dynamics of the coupled kicked rotors. Due to the boundedness of the Floquet quasi-energy spectrum there exists limitations on the interacting lattice models that can be mapped to quantum kicked rotors; in particular, no extensive observable can be mapped in the thermodynamic limit.
\end{abstract}

\maketitle

\sect{Introduction}
It was observed by Anderson~\cite{PhysRev.109.1492} that in one-dimensional systems an uncorrelated random potential induces destructive self-interference in such a way as to exponentially localise all non-interacting eigenstates of a quantum Hamiltonian, even with infinitesimally weak disorder. As a consequence, energy injected into the system resonantly with a trapped mode cannot result in any transport phenomena. In the original Anderson model, diffusion of electrons is suppressed by coherent back-scattering from magnetic impurities, but the one-dimensional localisation phenomenology is universal to wave mechanics~\cite{0034-4885-56-12-001} and has since been widely observed in experiments with matter waves~\cite{Billy2008,Roati2008}, light waves~\cite{PhysRevLett.100.013906,PhysRevLett.96.063904,Wiersma1997}, microwaves~\cite{Dalichaouch1991,Chabanov2000} and sound waves~\cite{WEAVER1990129} alike. The physics is strongly dependent on the dimensionality~\cite{PhysRevLett.42.673, Schwartz2007}; in particular, if $d \geq 3$ one can expect a mobility edge~\cite{Bulka1985}, an energy separating localised and extended eigenstates.

While sufficient, disorder is not a necessary condition for single-particle localisation. Indeed, the disorder need not exist directly in configuration space as an external potential, and localisation can also be induced by classically chaotic dynamics in momentum space. The resulting quantum suppression of diffusion in a system that is classically chaotic, termed `dynamical localisation'~\cite{casati1979stochastic,Izrailev1980}, was experimentally first observed with thermal non-condensed ultra-cold atoms realizing an atomic kicked rotor~\cite{PhysRevLett.73.2974}. A realization of the chaotic quasi-periodic kicked rotor with three incommensurate frequencies has permitted an unambigous observation of the metal-insulator Anderson transition in $d = 3$~\cite{PhysRevA.80.043626}. The model of the quantum kicked rotor~\cite{IZRAILEV1990299} has enjoyed several decades of intense research efforts as a paradigm quantum system that is chaotic in its classical limit to study the appearance of classical dynamical chaos in quantum mechanics~\cite{stockmann2006quantum,CHIRIKOV1979263,doi:10.1063/1.524170}. More recently, quantum chaos continues to play an important role in understanding experimental results of quantum transport in mesoscopic semiconductor devices~\cite{nakamura2004quantum,scholl2013theory}.

Similarly, Aubry and Andr\'e (AA) observed~\cite{aubry1980analyticity} a localisation transition not dissimilar from the mobility edge in a one-dimensional single-particle tight-binding model~\cite{PhysRevLett.82.4062} where the Anderson disorder potential is replaced by a quasi-periodic potential with a period that is incommensurate with the underlying lattice. The transition occurs as a function of the strength of the potential energy separating a phase where all single-particle eigenstates are localized (strong potential) from a phase where all eigenstates are extended (weak potential)~\cite{0305-4470-16-9-015}. At the transition point the eigenstates are multifractal~\cite{0295-5075-4-5-014,0305-4470-25-20-005}. The AA model has been experimentally realized with matter~\cite{PhysRevLett.106.230403} and light waves~\cite{PhysRevLett.103.013901}. Soon after AA had published their model, Fishman \textit{et al.} showed that the kicked rotor is related to Anderson's original problem of motion of a quantum particle in a one-dimensional tight-binding lattice in the presence of a quasi-periodic AA potential~\cite{PhysRevLett.49.509,PhysRevA.29.1639}. Their results showed for the first time that the dynamical localisation in momentum space of the kicked rotor model can be understood in terms of Anderson localisation in real space. Despite differences in the character of uncorrelated and correlated (quasi-periodic) randomness between the two models, the wave functions and the nature of the spectra are essentially the same in the two pictures. 

In parallel, efforts toward understanding the effect of interactions on both dynamical localisation and Anderson localisation have received significant attention~\cite{PhysRevB.21.2366, 0022-3719-14-3-007,PhysRevLett.70.1787,PhysRevLett.102.024101,PhysRevLett.61.659,0253-6102-51-3-17}. It was generally believed that localisation does not survive under interactions due to loss of coherence and dephasing, which generally collapse wave interference effects.
However, in 2006, Basko, Aleiner, and Altshuler showed that under suitable conditions interactions are insufficient to thermalise the system and localisation can in fact persist in a disordered quantum many-body system~\cite{BASKO20061126,ivanov2007problems}. This effect, termed `many-body localisation' (MBL), is destroyed when interactions become sufficiently strong leading to a dynamical phase transition to an `ergodic' conductive phase which supports quantum transport of energy~\cite{doi:10.1146/annurev-conmatphys-031214-014726}. Ergodic phases correspond to thermal systems, which can be described by the established theory of quantum statistical mechanics: at long enough times every sufficiently small subsystem reaches an effectively thermal Gibbs state~\cite{doi:10.1002/andp.201700169}. The MBL phase, in contrast, cannot be captured in terms of the ensembles of quantum statistical mechanics because it is non-erdogic and it violates the eigenstate thermalisation hypothesis (ETH)~\cite{PhysRevB.75.155111}. Therefore, the MBL transition is not a quantum phase transition, and it cannot be observed in terms of thermodynamics. For modern review articles on the unusual properties of the MBL phase we refer to e.g. Refs.~\cite{doi:10.1002/andp.201700169,doi:10.1146/annurev-conmatphys-031214-014726}.

Recent numerical evidence has shown that, similarly to the case of Anderson's model, uncorrelated disorder is not needed for localisation, and a quasi-periodic AA potential can also support the MBL phase~\cite{PhysRevB.87.134202,2018arXiv181101912M}. A natural question arises: Can MBL in an interacting extension of the AA model be mapped to the physics of coupled quantum kicked rotors in the same way as Anderson localisation in the AA model can be mapped to dynamical localisation of a single kicked rotor? As our key result, we show here that the answer appears to be affirmative. By analogy with the single-particle case, we term the new phase of the coupled kicked rotors `many-body dynamical localisation' (MBDL). However, in the Floquet theory of quantum kicked rotors energy is only defined as a quasi-energy upto multiples of $2\pi \hbar /T$, where $T$ is the kicking period. As we will show below this imposes important limitations on the types of lattice models that can mapped to quantum kicked rotors. 

The first indication of preservation of dynamical localisation in the presence of interactions for coupled kicked rotors was reported by Toloui and Ballentine~\cite{2009arXiv0903.4632T}, contrasting earlier predictions of the delocalising effect of interactions in configuration space~\cite{PhysRevLett.61.659,0253-6102-51-3-17}. Further support for dynamical many-body localisation was recently reported in an integrable linear kicked-rotor system~\cite{PhysRevB.94.085120}, which was later generalised for upto three coupled kicked rotors in a non-integrable relativistic model~\cite{PhysRevB.95.064303}. Qin \textit{et al.} showed that contact interactions in configuration space preserve dynamical localization for the center-of-mass momentum, but destroy it for the relative momentum for any non-zero strength of the interaction~\cite{Qin2017}. However, understanding more generally the existence of the many-body dynamical localisation effect remains unclear, especially for more than 2 to 3 rotors and under coupling that occurs in momentum space instead of the more commonly taken configuration space. Here, we establish analytically a mapping between an interacting extension of the Aubry-Andr\'e tight-binding model of $N$ particles supporting MBL and suitably coupled $N$ quantum kicked rotors in momentum space. 

\sect{Lattice and kicked-rotor models}
We start by considering the following fermionic lattice model~\cite{PhysRevB.87.134202}:
\begin{equation}
\label{eqn:HMBL}
\begin{split}
\hat{H} &= \sum_{j=0}^{L-1}\left[h_j \hat{n}_j + J(\hat{c}^\dagger_j \hat{c}_{j+1}  + \hat{c}^\dagger_{j+1} \hat{c}_j) + V \hat{n}_j \hat{n}_{j+1}  \right] \\
&\equiv  \hat{H}_0 + V\hat{H}_1.
\end{split}
\end{equation}
Here, $\hat{c}_j$ annihilates a fermion~\footnote{In $d = 1$, the difference in the Hamiltonian~\eqref{eqn:HMBL} between hard-core bosons and fermions is apparent only through the boundary conditions~\cite{PhysRevB.87.134202}. With open boundary conditions, the Hamiltonian is identical for both fermions and bosons.} at site $j$, and $\hat{n}_j = \hat{c}_j^\dagger \hat{c}_j \in \{0,1\}$ is the particle number operator at site $j$. The three terms in the Hamiltonian~\eqref{eqn:HMBL} correspond to an onsite potential $h_j$, nearest-neighbor hopping $J$, and nearest-neighbor interaction $V$ respectively. When $V = 0$ and $h_j$ is either truly random, or the quasi-periodic Aubry-Andr\'e function $h_j = h \cos{(2\pi \alpha j + h_0 )}$ with $\alpha$ irrational and an arbitrary offset $h_0$, we recover Anderson localisation. In this case, how to connect the non-interacting model~\eqref{eqn:HMBL} with a single quantum kicked rotor is shown in Refs.~\cite{PhysRevLett.49.509,PhysRevA.29.1639}. The related case of a single particle hopping in an $N$-dimensional potential $\sum_{i = 1}^N h_{l_i}$ has been mapped to $N$  quantum kicked rotors coupled at the kicks in Ref.~\cite{PhysRevE.97.022202}; however, we do not consider coupling at the kicks here.

The number-state basis $\{ \ket{n_0, n_1, \ldots, n_{L-1}} \}$, where $n_i = \{0,1\}$ is the number of particles at site $i$, constitutes for a tight-binding ansatz: $\lket = \sum_\n c_{\n} \ket{\n}$, where we defined the basis vector $\n= (n_0, n_1, \ldots, n_{L-1})$. Substitution into the many-body Schr\"odinger equation $\hat{H} \lket = E \lket$ and projection from the left on to $\bra{\m}$ gives the general equation 
\begin{equation}
\label{eqn:many-body_Sch}
c_\m \braket{\m | \hat{H} | \m} + \sum_{\n \neq \m} c_\n  \braket{\m | \hat{H} | \n} = E c_\m.
\end{equation}
In the one-particle case, we take $\bra{\m} = \bra{\m_i} = \bra{0,\ldots,0,1,0,\ldots,0}$, where the 1 is at the $i$th site. Since $\sum_{\n \neq \m_i} c_\n  \braket{\m_i | \hat{H}_0 | \n} = J\left( c_{\m_{i-1}} + c_{\m_{i+1}}  \right) $ and $\sum_{\n \neq \m_i} c_\n   \braket{\m_i | V \hat{H}_1 | \n}  = 0$, where $\m_k$ is zero except for 1 at the $k$th site, we recover the expected single-particle result $h_i c_{\m_i}   + J\left( c_{\m_{i-1}} + c_{\m_{i+1}}  \right) = E c_{\m_i}$.

Let us now take $\bra{\m} \equiv \bra{(l_1,l_2,\ldots,l_N)}$ with every site empty except for the sites $l_1<l_2<\ldots<l_N$, for which
\begin{subequations}
\begin{align}
\label{eqn:TBN1-1}
\braket{\m | \hat{H} | \m} &= \sum_{i = 1}^N h_{l_i}+ V\sum_{i = 1,j>i}^{N} \delta_{|l_i-l_{j}|,1},\\
\label{eqn:TBN1-3}
 \sum_{\n \neq \m} c_\n   \braket{\m| V \hat{H}_1 | \n}  &= 0.
\end{align}
\end{subequations}
The result is the addition of contact interactions of strength $V$ in the $N$-particle configuration space. For $N = 2$, we have
\begin{equation}
\label{eqn:TBN2}
\begin{split}
&J\left( c_{(l_1-1,l_2-1)} + c_{(l_1+1,l_2+1)} + c_{(l_1-1,l_2+1)} + c_{(l_1+1,l_2-1)} \right) \\
&\qquad + c_{(l_1,l_2)}  \left( h_{l_1} + h_{l_2}+ V \delta_{|l_1-l_2|,1} \right) = E c_{(l_1,l_2)}.
\end{split}
\end{equation}
When $V = 0$, this is equivalent to a single particle hopping in a $d=2$ potential $h_{l_1} + h_{l_2}$.

Our goal is to map the interacting lattice model~\eqref{eqn:many-body_Sch} of $N$ particles to quantum motion of $N$ coupled kicked rotors governed by the Floquet Hamiltonian
\begin{equation}
\label{eqn:HF}
\HF(t) = K(\hat{\mathbf{L}}) + \Vt \sum_n \delta(t - nT),
\end{equation}
where $K(\hat{\mathbf{L}}) \equiv \K$ and $\Vt$ are functions of angular momentum and angle, respectively, to be determined later. The operator $\Vt$ depends on $N$ angles through the vector $\hat{ \boldsymbol{\theta}} = (\hat{\theta}_1, \hat{\theta}_2, \ldots, \hat{\theta}_N)^\mathrm{T}$, and $\K$ depends on $N$ angular momenta $ \boldsymbol{\hat{L}} = (\hat{L}_1, \hat{L}_2, \ldots, \hat{L}_N)^\mathrm{T}$. The Hamiltonian is periodic in time, $\HF(t) = \HF(t + T)$, and the corresponding Floquet operator reads
\begin{equation}
\label{eqn:FloqOp}
\F = \rme^{-\frac{\rmi}{\hbar} \Vt} \rme^{-\frac{\rmi}{\hbar} \K T}.
\end{equation}
In general, knowledge of the propagator $\F$ over the fundamental period $T$ contains all the necessary information to generate the discrete quantum map $\F(nT) = \F^n$ and study the long-time dynamics of periodically driven quantum systems in a stroboscopic manner. The eigenstates of the angular momentum operator, $\hat{L} =- \rmi \hbar \frac{\partial}{\partial \theta}$, are the rotational states $\braket{\theta|\ell} = \frac{1}{\sqrt{2\pi}} \rme^{\rmi \ell \theta}$ enumerated by an integer quantum number $\ell \in \mathbb{Z}$ with eigenvalue $\hbar \ell$. For a single quantum kicked rotor $K(\hat{L}) = \frac{\hat{L}^2}{2I}$ and $V(\hat{\theta}) = k \cos{(\hat{\theta})}$, where $k$ is the strength of the kicks. We will show in what follows how to generalise $K(\hat{L})$ and $V(\hat{\theta})$ to establish the mapping to the MBL model.

\sect{Mapping between the models}
We will now establish a mapping between the Hamiltonians $\HF$ and $\hat{H}$, Eqs.~\eqref{eqn:HF}  and~\eqref{eqn:HMBL} respectively. Following Fishman \textit{et al.}~\cite{PhysRevLett.49.509,PhysRevA.29.1639}, let $\ket{A}$ be the eigenfunction of the Floquet operator~\eqref{eqn:FloqOp} with eigenphase $\phi$, that is, $\F\ket{A} = \rme^{-\rmi \phi}\ket{A}$. The Floquet state $\ket{A}$ is then an eigenstate of the stroboscopic dynamics. The eigenvalue equation can be combined with its time-reversed dual $\F^\dagger\ket{A} = \rme^{\rmi \phi}\ket{A}$ to read
\begin{equation}
\label{eqn:FloqEEq-Utf0}
\rme^{\mp\rmi \K T/(2\hbar)} \rme^{\mp \frac{\rmi}{\hbar} \Vt} \rme^{\mp \rmi \K T/(2\hbar)}\ket{\tilde{A}} = \rme^{\mp\rmi \phi}  \ket{\tilde{A}},
\end{equation}
where $\ket{\tilde{A}} =  \rme^{\rmi\K T/(2\hbar)}\ket{A}$. Let us define the local angular momentum eigenstates $\ket{\tn} = \ket{\tilde{n}_1, \tilde{n}_2, \ldots, \tilde{n}_{N}}$ with 
$\tilde{\n} \in \mathbb{Z}^N$, such that the free propagator $\K = \frac{1}{2I} \sum_{i=1}^N \hat{L}^2_i $ with $\hat{L}_i = -\rmi \hbar \partial_{\theta_i}$ the angular momentum operators with eigenvalues $\hbar n_i$ and moment of inertia $I$ acts as $\frac{ \K T}{2 \hbar} \ket{\tn} = \varphi(\tn) \ket{\tn}$ where $\varphi(\tn) \equiv \frac{\kbar}{4}   \sum_{i=1}^N\tilde{n}_i^2 \in \mathbb{R}$ and $\kbar \equiv \hbar T / I$.

We now generalise the kinetic energy operator $\K \to \kap = \K + \K^\prime$, where $\K^\prime$ is undetermined at the moment apart from the following requirements: (i) $[\K, \K^\prime ] = 0$ allowing us to factor the operator $\rme^{\rmi \kap T/(2\hbar)}$ into two exponentials; and (ii) In accordance with Eq.~\eqref{eqn:TBN1-1} $\K^\prime$ represents nearest-neighbour contact interactions in angular momentum space, diagonal in the Fock space. We define the function $\varphi^\prime(\tn)$ through the action of the new operator: $\K^\prime\ket{\tn} = \frac{2\hbar^2}{I \kbar}  \varphi^\prime(\tn) \ket{\tn}$. It is then convenient to define the function $\Phi(\tn) \equiv \varphi(\tn) + \varphi^\prime(\tn)$. Inserting resolution of the unity $\mathbf{1} = \sum_\tn \ket{\tn}\bra{\tn}$ and projecting from the left onto $\bra{\tm}$, we find from Eq.~\eqref{eqn:FloqEEq-Utf0}:
\begin{equation}
\begin{split}
\label{eqn:FloqEEq-Utf-TB-pre1-s}
&\left(-\rme^{\pm\rmi \phi}  + \braket{\tm|\F_\pm |\tm}\right) \braket{\tm|\tilde{A}} \\
&\qquad +  \sum_{\tn \neq \tm}\rme^{\rmi \left[\Phi(\tm) - \Phi(\tn) \right]} \braket{\tm|\F_\pm |\tn}\braket{\tn|\tilde{A}}= 0,
\end{split}
\end{equation}
where $\F_+ = \F^\dagger$, $F_- = \F$. We can evaluate directly to obtain $\braket{\tm| \F| \tn} =  \rme^{-2\rmi \, \Phi(\tm)}  \braket{\tm| \rme^{-\frac{\rmi}{\hbar} V(\hat{\boldsymbol{\theta}})} | \tn}$ and $\braket{\tm| \F^\dagger | \tn} = \rme^{2\rmi \, \Phi(\tn)}  \braket{\tm| \rme^{\frac{\rmi}{\hbar} V(\hat{\boldsymbol{\theta}})} | \tn}$. Using the time-reversal symmetry of $\HF$, fixing the Floquet gauge without loss of generality, labeling the on-site potential by $h_\tm $, and the wavefunction $c_\tm \equiv \braket{\tm|\tilde{A}} $, we obtain from the system~\eqref{eqn:FloqEEq-Utf-TB-pre1-s}~\cite{PhysRevE.97.022202}
\begin{equation}
\label{eqn:FloqEEq-Utf-TB}
h_\tm c_\tm + \sum_{\tn \neq \tm} W_{\tm,\tn} c_\tn = \cos{\left(\phi \right)} c_\tm ,
\end{equation}
where
\begin{equation}
\label{eqn:FloqEEq-Utf-TB-pre4}
W_{\tm,\tn} = \mathrm{Re}\left\lbrace  \rme^{\rmi \left[\Phi(\tm) + \Phi(\tn) \right]}  \int_0^{2\pi} \frac{\d^N \boldsymbol{\theta}}{(2\pi)^N} \rme^{\frac{\rmi}{\hbar} \Vts} \rme^{-\rmi (\tm - \tn)\cdot \boldsymbol{\theta}} \right\rbrace
\end{equation}
and $h_\tm \equiv  W_{\tm,\tm}$. We now focus on the tight-binding dynamics described by Eq.~\eqref{eqn:FloqEEq-Utf-TB}, which corresponds to an $N$-body Schr\"odinger equation with a local potential $h_\tm $ and energy eigenvalue $\cos{\left(\phi\right)}$. While not dissimilar from the map established in Ref.~\cite{PhysRevE.97.022202}, it is important to notice that Eq.~\eqref{eqn:FloqEEq-Utf-TB} is distinct in that we also consider interactions in the lattice model.

To produce lattice models similar to Eq.~\eqref{eqn:TBN1-1}, we identify $c_\tm = c_{(l_1,l_2,\ldots, l_N)}$, label $h_\tm \equiv  \epsilon(\tm) + V(\tm)$, and take 
\begin{equation}
\begin{split}
\label{eqn:FloqEEq-Utf-TB-map-1}
 &\int_0^{2\pi} \frac{\d^N \boldsymbol{\theta}}{(2\pi)^N} \cos{  \left[ \frac{\Vts}{\hbar} + 2\varphi(\tm) + 2  \varphi^\prime(\tm)  \right]}  \\
 &\qquad  =\epsilon(\tm) +   V(\tm).
\end{split}
\end{equation}
We will later specialise to the case $V(\tm) = V  \sum_{i = 1,j>i}^{N} \delta_{|\tilde{m}_i-\tilde{m}_j|,1}$ to connect with the model~\eqref{eqn:HMBL}, but for the moment we keep the formalism general. Since the quasi-energy is defined only modulo $2\pi \hbar/T$, the left hand side of Eq.~\eqref{eqn:FloqEEq-Utf-TB-map-1} is bounded, which means that the energies and lattice models we can map to must be bounded as well. This is a fundamental restriction with the Floquet Hamiltonian, and it has important consequences for the thermodynamic limit as we describe below.

The purpose of the  interaction operator $\K^\prime$ becomes clear: it can be chosen suitably to engineer desired interactions in the lattice model. In particular, when $V(\tm) = 0$, we can set $ \varphi^\prime(\tm)  = 0$ in which case Eq.~\eqref{eqn:FloqEEq-Utf-TB-map-1} reduces to
\begin{equation}
\label{eqn:FloqEEq-Utf-TB-map-2b}
\epsilon(\tm)  = \int_0^{2\pi} \frac{\d^N \boldsymbol{\theta}}{(2\pi)^N} \cos{ \left[ \frac{\Vts}{\hbar} + 2 \varphi(\tm)\right]},
\end{equation}
which is a quasi-periodic on-site potential e.g. if $\Vts =  \sum_{l = 1}^N  k^{(l)} \cos{\left(\theta_l \right)}$. When $\kbar \mod 4\pi = 0$, the potential~\eqref{eqn:FloqEEq-Utf-TB-map-2b} becomes a constant and cannot induce any localisation -- this case corresponds to the quantum resonances of the kicked rotor~\cite{IZRAILEV1990299}. The same resonances occur in the original single-particle map reported in Refs.~\cite{PhysRevLett.49.509,PhysRevA.29.1639}. 
Using Eq.~\eqref{eqn:FloqEEq-Utf-TB-map-2b} in Eq.~\eqref{eqn:FloqEEq-Utf-TB-map-1}, we find the equation
\begin{equation}
\label{eqn:FloqEEq-Utf-TB-map-5}
  \cos{\left[  2   \varphi^\prime(\tm) \right]} + 
\Gamma_1(\tm)  \sin{\left[  2   \varphi^\prime(\tm) \right]}   
= \Gamma_2(\tm),
\end{equation}
where
\begin{subequations}
\begin{align}
\Gamma_1 &= - \frac{ \int_0^{2\pi} \frac{\d^N \boldsymbol{\theta}}{(2\pi)^N} \sin{  \left[ \frac{\Vts}{\hbar} + 2\varphi(\tm) \right]}}{\int_0^{2\pi} \frac{\d^N \boldsymbol{\theta}}{(2\pi)^N} \cos{  \left[ \frac{\Vts}{\hbar} + 2\varphi(\tm)   \right]}  },\\
\Gamma_2 &= 1 + \frac{V(\tm)}{ \int_0^{2\pi} \frac{\d^N \boldsymbol{\theta}}{(2\pi)^N} \cos{  \left[ \frac{\Vts}{\hbar} + 2\varphi(\tm) \right]} },
\end{align}
\end{subequations}
which must be solved for $ \varphi^\prime(\tm)$ once the form of $\Vts$ and $V(\tm)$ has been chosen. We have the analytic solution
\begin{equation}
\label{eqn:FloqEEq-Utf-TB-map-50} 
2\varphi^\prime(\tm) = \arctan{\left[ \frac{\pm \Gamma_1 \gamma + \Gamma_2   }{\Gamma_1 \Gamma_2 \mp \gamma} \right]} \in [-\pi,\pi],
\end{equation}
where $\gamma \equiv \sqrt{1 + \Gamma_1^2 - \Gamma_2^2}$. The general map~\eqref{eqn:FloqEEq-Utf-TB-map-50} is the key result of this work.

\begin{figure}[t]
  \centering
        \begin{tikzpicture}
        \def\x{4.5};        \def\y{3.0};
        \def\v{-0.9};      \def\w{10.0};	     \def\u{10.0};
              \node at (0,0) {    \includegraphics[width=0.23\textwidth,angle=0]{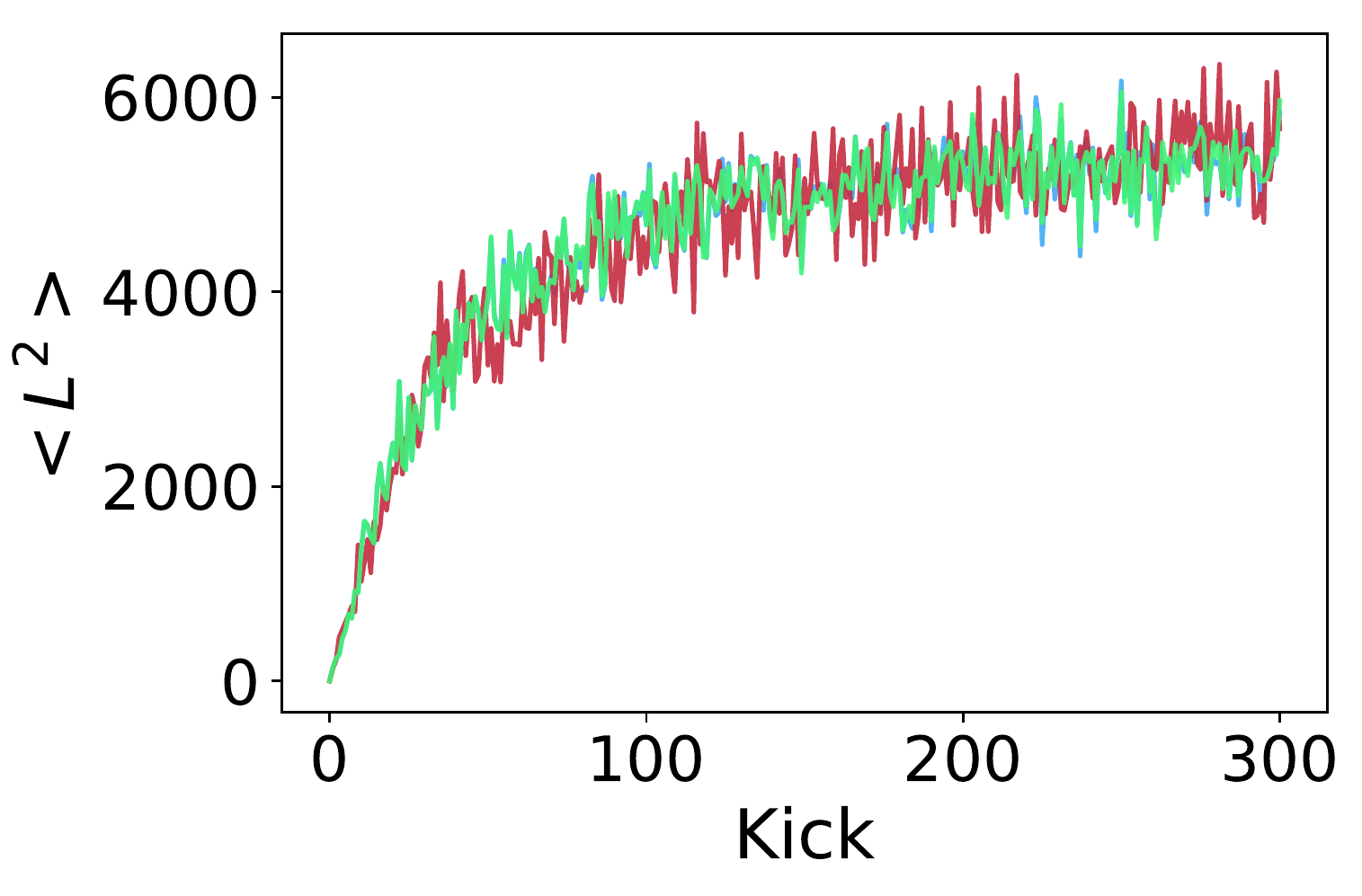}};
              \node at (\x,0) {    \includegraphics[width=0.23\textwidth,angle=0]{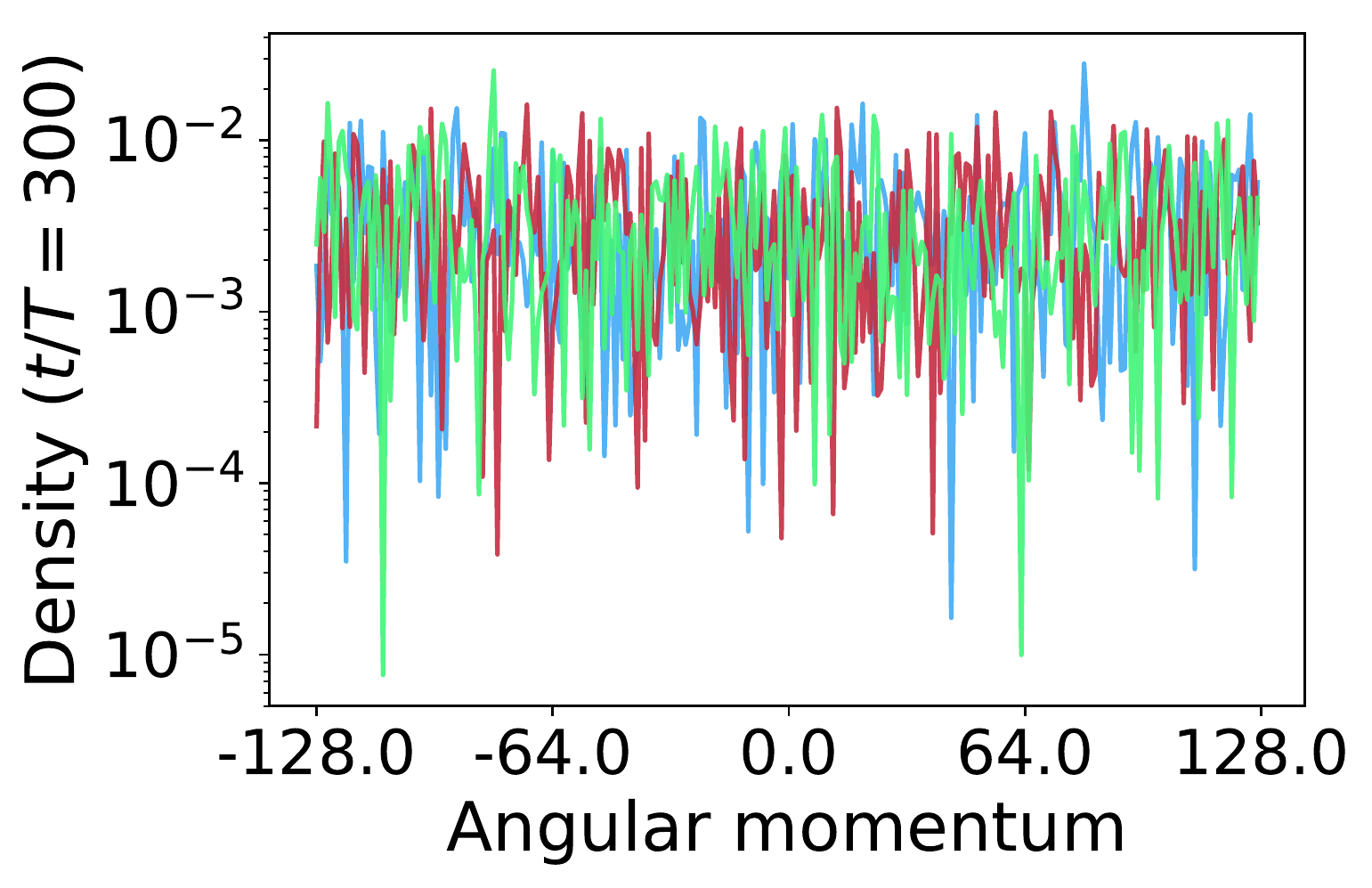}};
			 \node at (0,\y) {    \includegraphics[width=0.23\textwidth,angle=0]{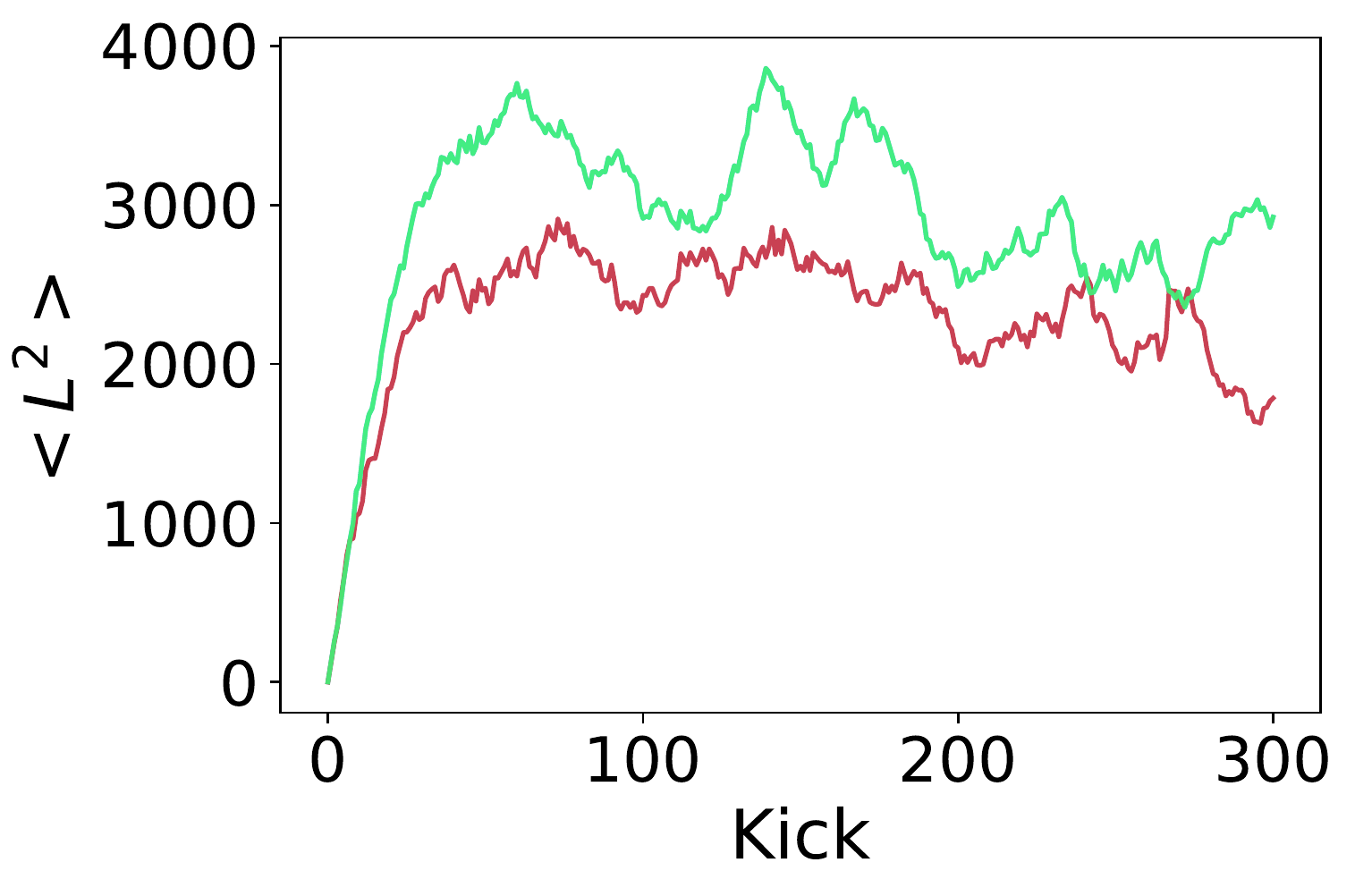}};
              \node at (\x,\y) {    \includegraphics[width=0.23\textwidth,angle=0]{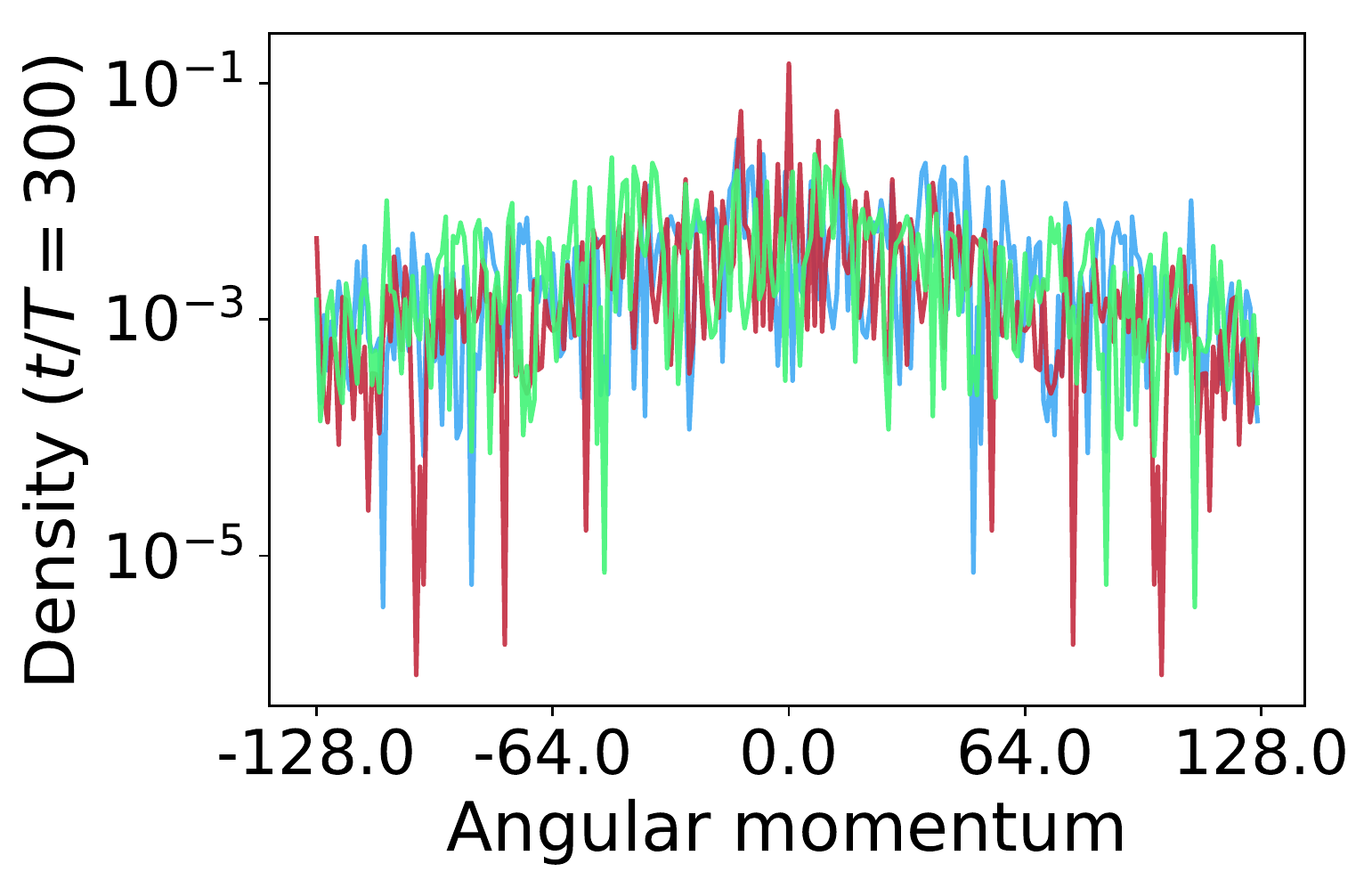}};
			 \node at (0,2*\y) {    \includegraphics[width=0.23\textwidth,angle=0]{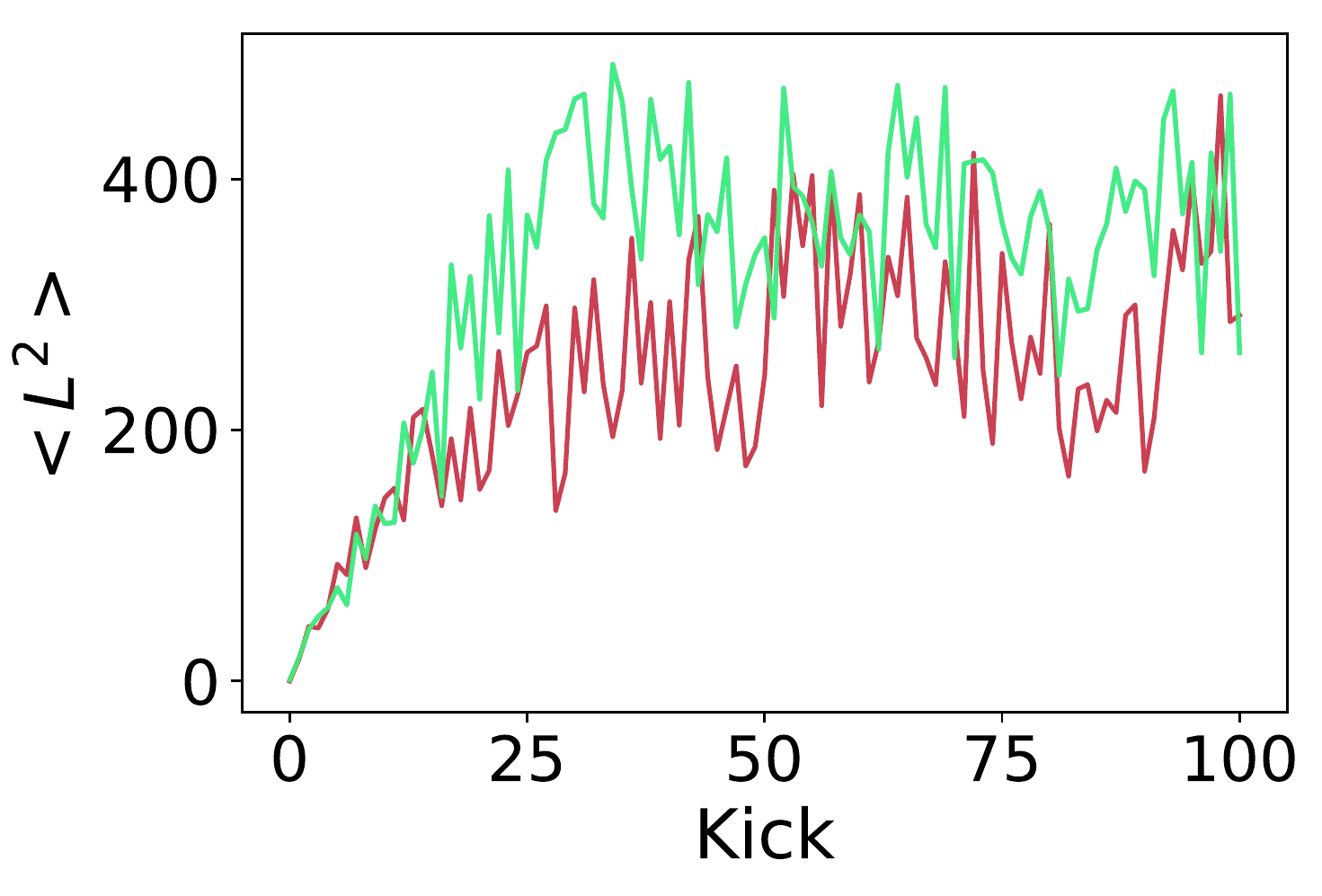}};
              \node at (\x,2*\y) {    \includegraphics[width=0.23\textwidth,angle=0]{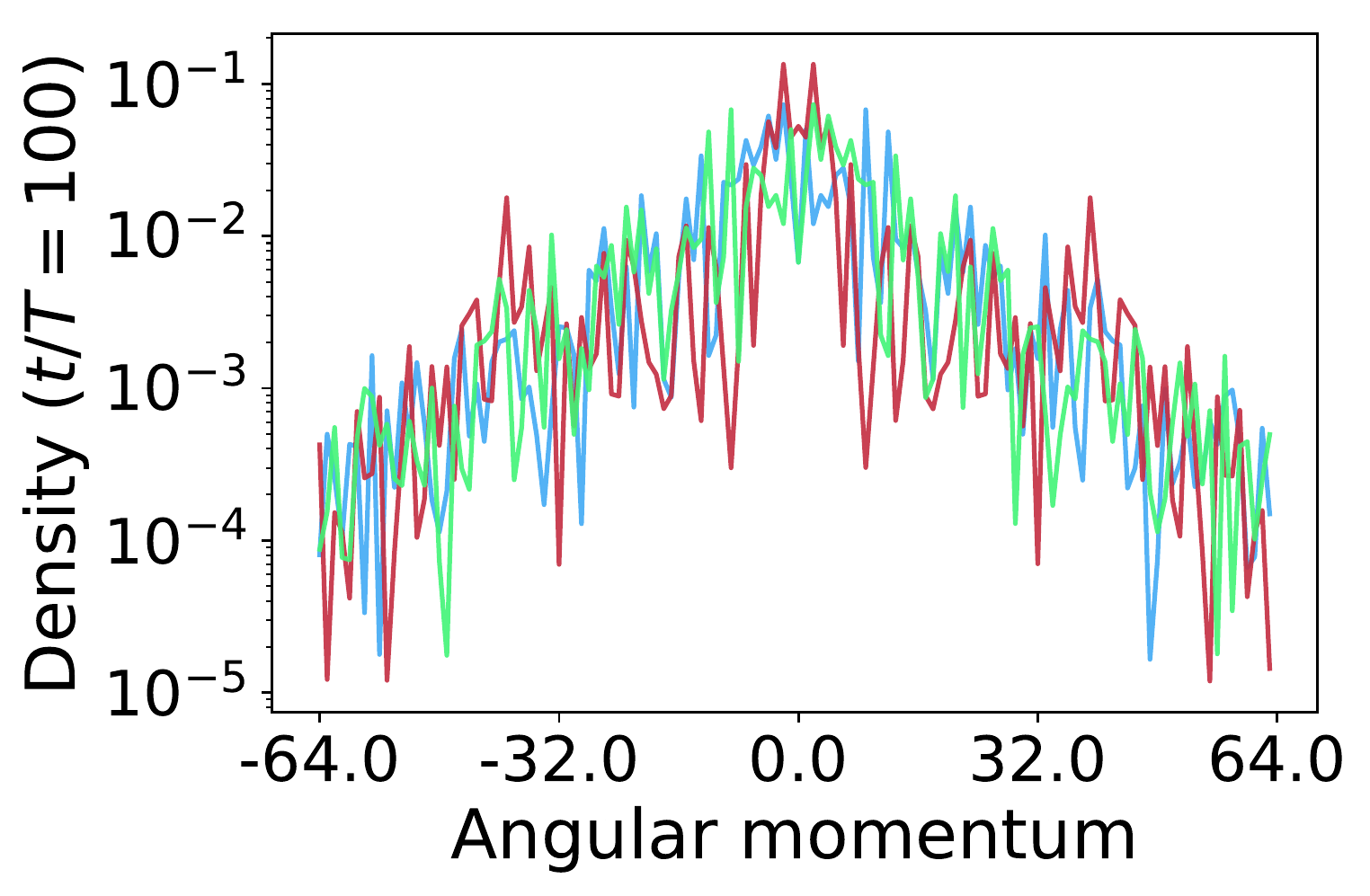}};
		     \node at (0,3*\y) {    \includegraphics[width=0.23\textwidth,angle=0]{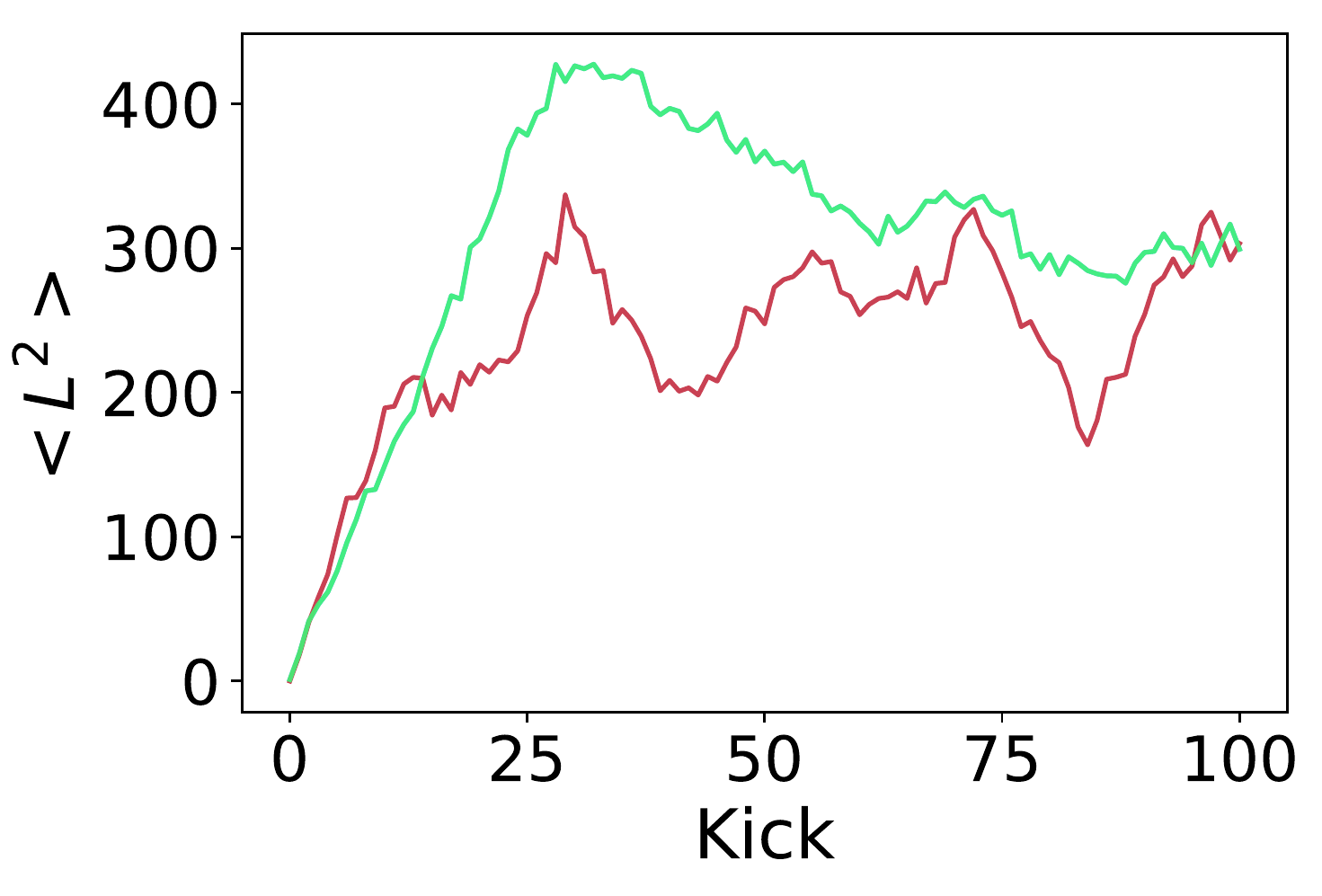}};
              \node at (\x,3*\y) {    \includegraphics[width=0.23\textwidth,angle=0]{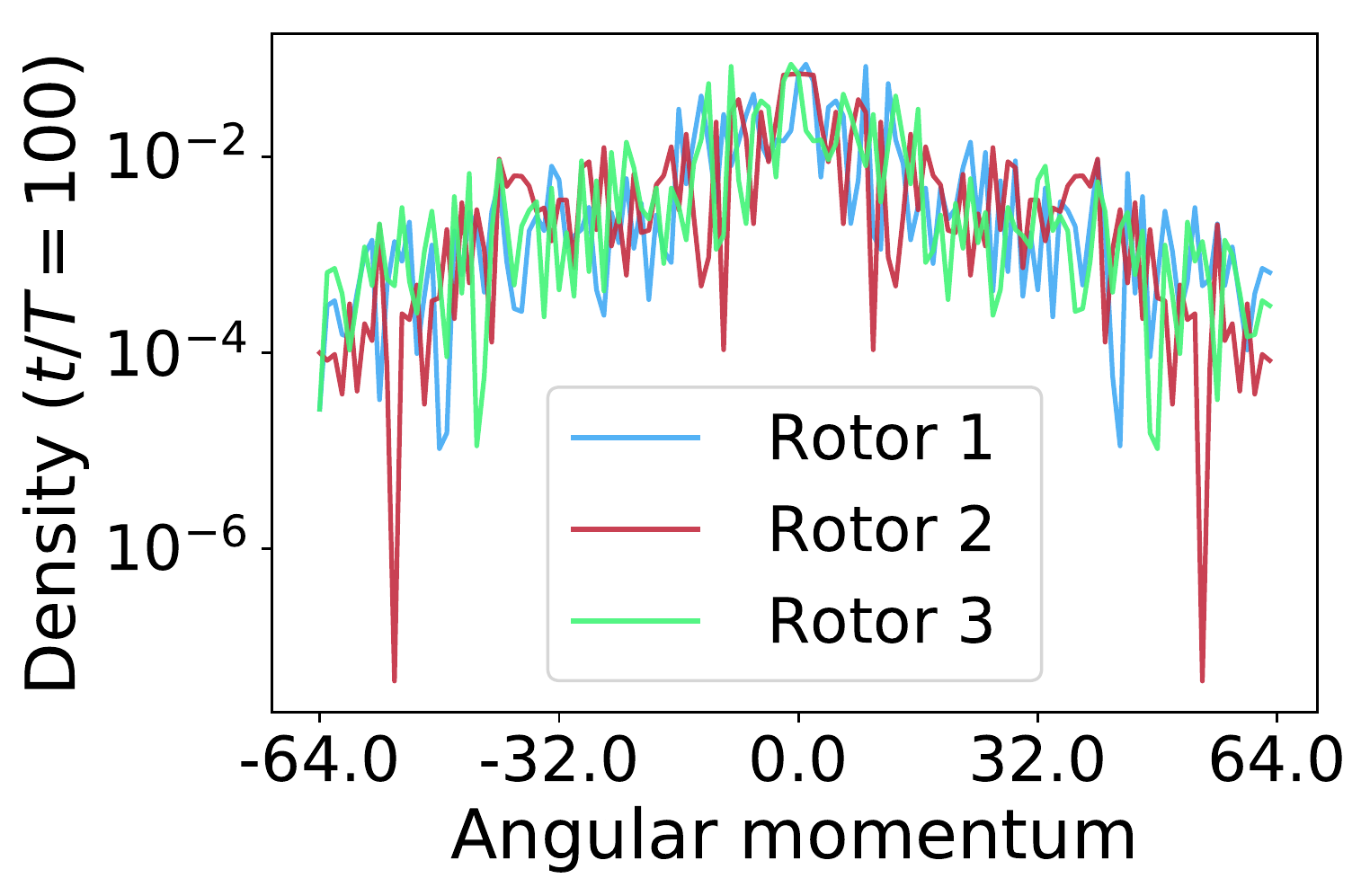}};

      \node at (\v,\w) {(a)};
            \node at (\v+\x,\u) {(b)};
                  \node at (\v,\w-\y) {(c)};
                        \node at (\v+\x,\u-\y) {(d)};
                        \node at (\v,\w-2*\y) {(e)};
                        \node at (\v+\x,\u-2*\y) {(f)};
                                              \node at (\v,\w-3*\y) {(g)};
                        \node at (\v+\x,\u-3*\y) {(h)};

                         \node at (\v+\x-0.1,\w-4*\y+0.1) {(i)};                       
              \node at (2.0,-\y-0.3) {    \includegraphics[width=0.2\textwidth,angle=0]{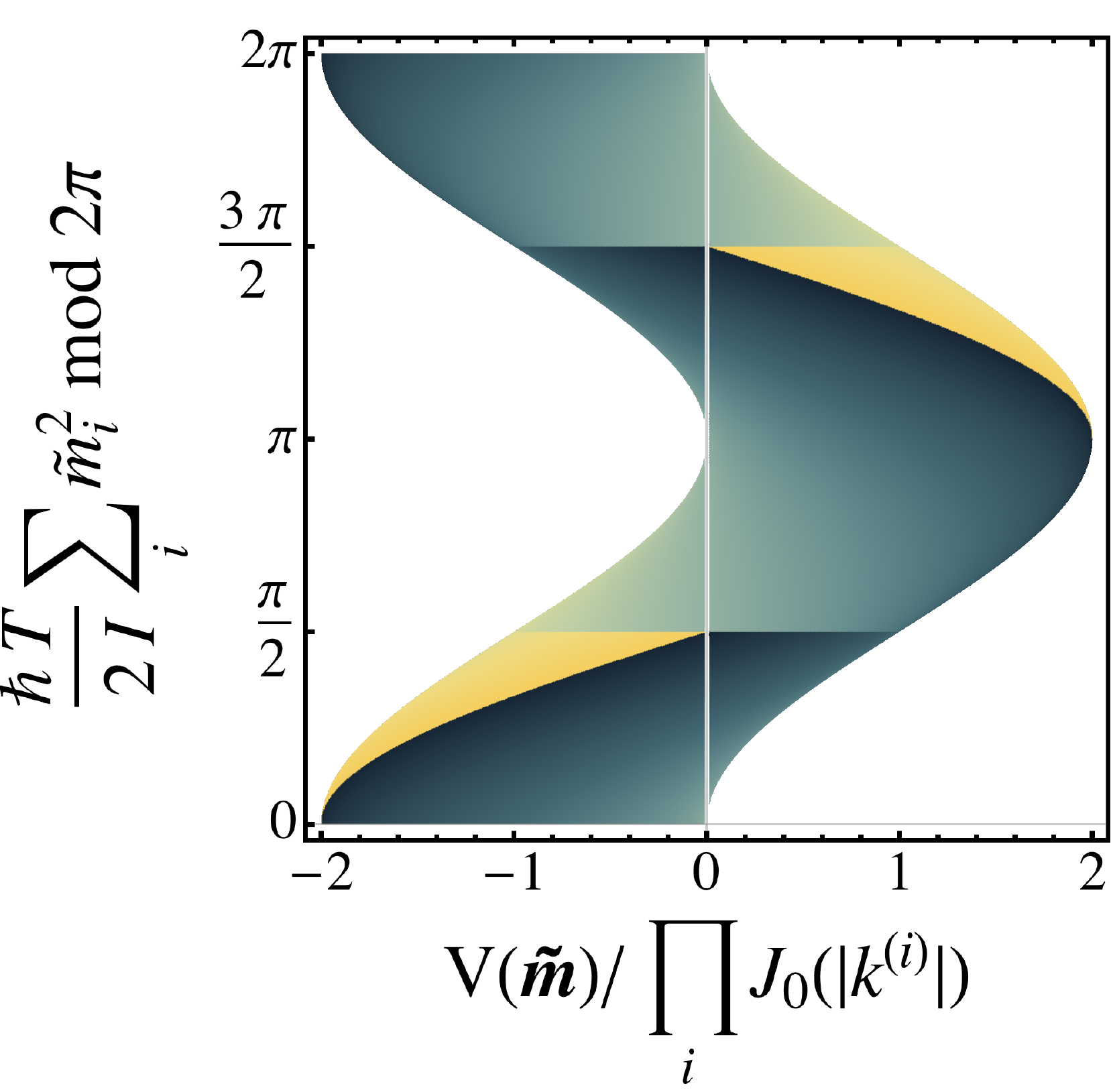}};
              \node at (2.0+2.2,-\y-0.05) {    \includegraphics[width=0.035\textwidth,angle=0]{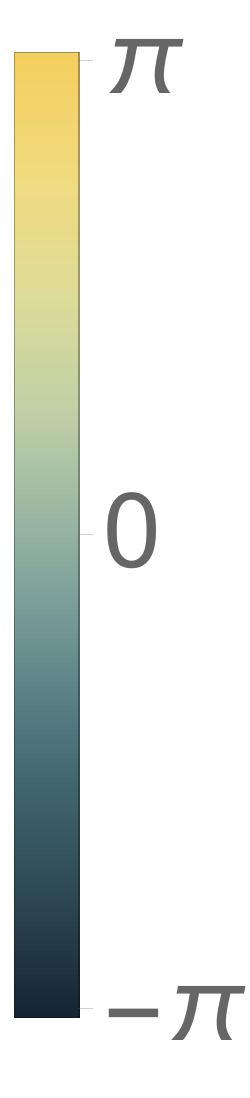}};

      \end{tikzpicture}
      
      \caption{ \label{fig:1} 
      Many-body dynamical localisation (MBDL) of three kicked rotors. 
      (a,b) Non-interacting case for weak kicking ($\Keff = 6.0$). 
      (c,d) With $\Keff = 6.0$ and interactions from solving Eq.~\eqref{eqn:FloqEEq-Utf-TB-map-5} with $V = 0.001$. Localisation persists and interactions change nothing qualitatively. 
      (e,f) Non-interacting case for strong kicking ($\Keff = 16.0$). 
      (g,h) With $\Keff = 16.0$ and interactions with $V = 0.001$. Localisation is either lost or significantly weaker, that is, of longer range. 
      (i) Real-valued solutions for $2\varphi^\prime(\tm)$. 
      In (a-h) $\Keff \equiv k^{(1)} = k^{(2)} = k^{(3)}$, $I = 0.1$, and $T  = \hbar = 1$. The initial condition is $\tilde{m}_1 = -1, \tilde{m}_2 = 0, \tilde{m}_3 = 1$.}   
\end{figure}
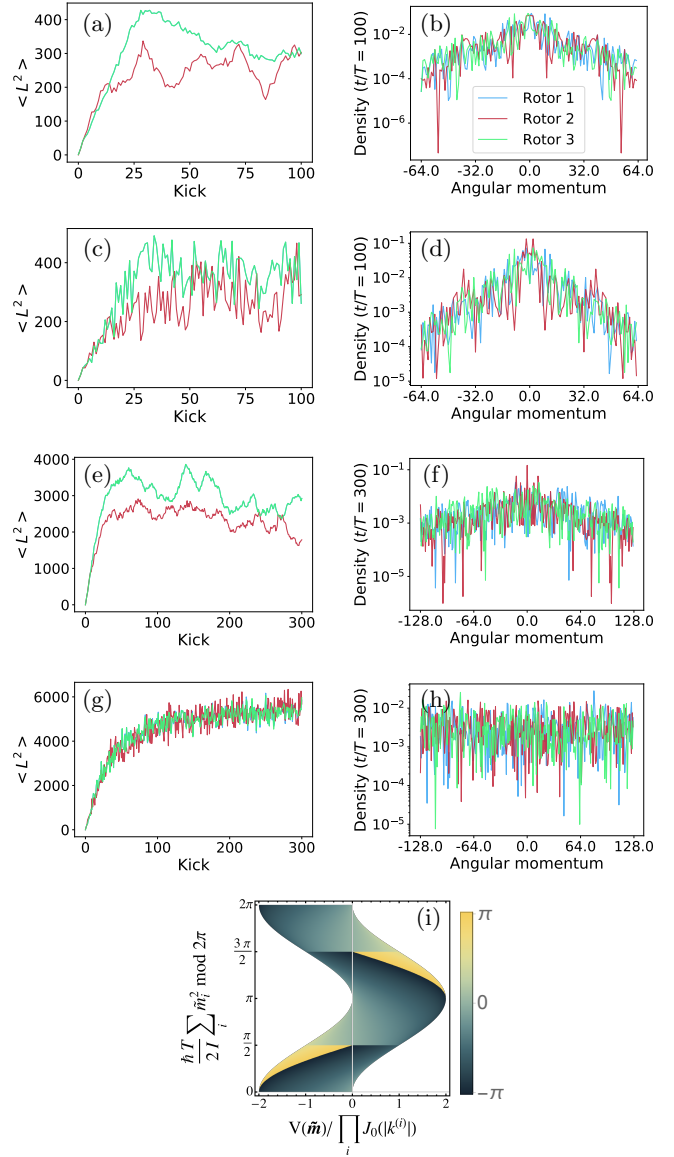

Let us choose $\Vts = \hbar \sum_{l = 1}^N  k^{(l)} \cos{\left(\theta_l \right)} $. Then $\Gamma_1 = \tan{\left[ 2\varphi(\tm)  \right]}$, and $\Gamma_2 = 1 +  V(\tm) / \left\lbrace  \cos{\left[ 2\varphi(\tm)  \right]} \prod_{l = 1}^N J_0 \left(|k^{(l)} | \right)\right\rbrace$. With this choice, the real-valued solutions to Eq.~\eqref{eqn:FloqEEq-Utf-TB-map-5} are shown in Fig.~\ref{fig:1}i. Notably, the solution (Fig.~\ref{fig:1}i) is identically zero when there are no interactions in the lattice model, that is, when $V(\tm)= 0$. Therefore, $\K^\prime$ assumes a similar form in the angular momentum many-body Fock space of the rotors as the interaction term is in the configuration space lattice model, here coupling only nearest-neighbours in angular momentum space. The white regions in Fig.~\ref{fig:1}i correspond to the eigenvalues of $\K^\prime$ becoming complex, and with those parameters we cannot map the Hermitian MBL model~\eqref{eqn:HMBL} to a Hermitian kicked rotor model. The white regions are essentially defined by the boundedness of the Floquet energy spectrum, which simultaneously imposes an upper bound for the magnitude of $V$ that we can consider in the map between the two models. Importantly, in the thermodynamic limit $N \to \infty$ the product $\prod_{l = 1}^N J_0 \left(|k^{(l)} | \right)$ vanishes, which enforces $V \to 0$ as well to keep the solution to Eq.~\eqref{eqn:FloqEEq-Utf-TB-map-5} real (Fig.~\ref{fig:1}i). Numerical integration gives the same result with configuration space interactions $\Vts \sim \cos{(\theta_i - \theta_j)}$. Therefore, it appears not possible to map an extensive quantity such as the contact interactions in the lattice model to an equivalent quantum kicked rotor model in the thermodynamic limit using our approach.

\sect{Numerical solution}
We now specialise to the case $V(\tm) = V  \sum_{i = 1,j>i}^{N} \delta_{|\tilde{m}_i-\tilde{m}_j|,1}$. For $N = 2$ we have $V(\tm)/V \in \{0,1\}$, and for $N = 3$ we have $V(\tm)/V \in \{0,1,2\}$. When $u \equiv V/h = 0$ and under periodic boundary conditions, the single-particle Aubry-Andr\'e model~\eqref{eqn:HMBL} has a metal-insulator transition at $g \equiv J / h = 0.5$ with all eigenstates extended (localised) for $g > 0.5$ ($g < 0.5$). This transition persists for finite $u$ at half-filling~\cite{PhysRevB.87.134202}, and has been observed with $N \geq 3$ kick-coupled rotors when $u = 0$ as a function of the hopping strength $|W_{\tm,\tn}|$, which increases with the kicking strength $k^{(l)}$~\cite{PhysRevE.97.022202}. In the context of the map~\eqref{eqn:FloqEEq-Utf-TB}, one can expect the mobility edge of the interacting lattice model~\eqref{eqn:HMBL} at finite $u \neq 0$ to separate a many-body dynamically localised (MBDL) phase from an extended ergodic phase of the kicked rotors. Both the strength of the kicks $k^{(l)}$ and interactions have the effect of making the wave packet spread faster, which corresponds to a larger effective hopping. Increasing the number of rotors, $N$, increases the effect of the interactions, but also imposes an upper bound for $V$. 

We have simulated numerically~\footnote{Our code can be found online at~\url{https://github.com/laantoi/open-science}} the quantum dynamics of coupled rotors described by $\HF$, where $\kap$ contains the coupling term $\K^\prime$ as described above (Fig.~\ref{fig:1}). Away from the quantum resonances, we observe for $N = 2$ and $N = 3$ what appears to be a metal-insulator transition as a function of the effective hopping, controlled via the kicking strength $\Keff \equiv k^{(1)} = k^{(2)} = k^{(3)}$. When $\Keff$ is small (Fig.~\ref{fig:1}a--d) coupling the rotors preserves dynamical localisation which corresponds to the MBDL regime. On the other hand, when $\Keff$ is sufficiently large, the coupled rotors delocalise over the entire truncated the Hilbert space (Fig.~\ref{fig:1}e--h). A defining feature of a mobility edge at energy $E_c$ is that the localisation length diverges, $\xi(E) \sim 1/(E - E_c)^\nu$. While suggestive of a mobility edge, our numerical results are by nature always insufficient to conclusively rule out exponential localisation with a larger localisation length that would become apparent by increasing the momentum cut-off $L$ that defines the Hilbert space truncation. More conclusive evidence may come from other localisation measures such as the participation ratio and level spacing statistics of the Floquet operator, which we will consider elsewhere. It is also important to note that while the map~\eqref{eqn:FloqEEq-Utf-TB-map-50} associates the tight-binding and kicked rotor models, and the numerical results indicate that dynamical localisation can persist under weak interactions, one should be careful when drawing deeper analogies. MBL is richer in phenomenology than simply Anderson localisation of $N$ particles, and it remains an interesting question how exactly these manifest through the map~\eqref{eqn:FloqEEq-Utf-TB-map-50}.

\sect{\label{sec:dc} Conclusions}
We have mapped the quantum motion of $N$ coupled kicked rotors to an interacting $N$-particle Anderson-Aubry-Andr\'e tight-binding problem supporting a many-body localised phase. With the contact interactions in angular momentum space considered here, we find numerical indications that the MBDL phase of the coupled kicked rotors persists for small values of the kicking strength, which corresponds to weak effective hopping in the lattice model.

\acknowledgements
This work was supported by the Austrian Academy of Sciences (P7050-029-011) and the Institute for Basic Science in Korea (IBS-R024-D1). LT would like to thank Kun-Woo Kim and Simone Notarnicola for helpful discussions.

\textit{Author contributions} -- 
LT designed and performed the research, and wrote the manuscript. LT and AA interpreted the results.

\bibliographystyle{apsrev4-1}
\bibliography{MBL}

\end{document}